\documentstyle[11pt]{article}
\newcommand{\beq}{\begin{equation}}
\newcommand{\eeq}{\end{equation}}
\newcommand{\beqr}{\begin{eqnarray}}
\newcommand{\eeqr}{\end{eqnarray}}

\newcommand{\sss}{\vspace{.2in}}

\begin{document}
\begin{titlepage}
\vspace{-.2in}\begin{flushright}\today\end{flushright}
\begin{center} {\Large {\bf  
Shape Invariance and Its Connection to Potential Algebra} }\\

\vspace*{1 in}

{\large Asim Gangopadhyaya$^{a,}$\footnote{e-mail: agangop@luc.edu, 
asim@uic.edu}, Jeffry V. Mallow $^{a,}$\footnote{e-mail: jmallow@luc.edu}
and Uday P. Sukhatme$^{b,}$\footnote{e-mail: sukhatme@uic.edu}.}
\end{center}
\begin{tabular}{l l}
$a$) &Department of Physics, Loyola University Chicago,
Chicago, IL 60626;\\
$b$) &Department of Physics, University of Illinois at
Chicago, (m/c 273) \\& 845 W. Taylor Street, Chicago, IL 60607-7059.\\
\end{tabular}
\vspace{1.0in}
\begin{center}
{\large {\bf Abstract}}\\
\end{center}

Exactly solvable potentials of nonrelativistic quantum mechanics are known to 
be shape invariant. For these potentials, eigenvalues and eigenvectors can be 
derived using well known methods of supersymmetric quantum mechanics.
The majority of these potentials have also been shown to possess a potential 
algebra, and hence are also solvable by group theoretical techniques. In this 
paper, for a subset of solvable problems, we establish a connection between 
the two methods and show that they are indeed equivalent.
\end{titlepage}
\newpage
\noindent
{\bf I. Introduction}\\

It is well known that most of the exactly solvable potentials of 
nonrelativistic quantum mechanics fall under the Natanzon 
class (\cite{Natanzon}) where the Schr\"{o}dinger equation reduces either 
to the hypergeometric or the confluent hypergeometric differential equations. 
A few exceptions are known (\cite{Spiridinov,Barclay}), where solvable
potentials are given as a series, and can not be written in closed form 
in general. With the exception of Ginnochio potential, all exactly solvable 
potentials are known to be shape invariant (\cite{Gendenshtein,Gendenshtein2}); 
i.e. their supersymmetric partners  are of the same shape, and their spectra 
can be determined entirely by an algebraic procedure, akin to that of the one 
dimensional harmonic oscillator, without ever referring to the underlying 
differential equations (\cite{Cooper}).  

Several of these exactly solvable systems are also known to possess what is 
generally referred to as a potential 
algebra (\cite{Alhassid,Barut,Englefield0,Englefield,Wu,Tangerman}). 
The Hamiltonian of these systems can be written as the Casimir of an underlying 
SO(2,1) algebra, and all the quantum states of these systems can be determined 
by group theoretical methods. 

Thus, there appear to be two seemingly independent algebraic methods for 
obtaining the complete spectrum of these Hamiltonians. In this paper, we 
analyze this ostensible coincidence. For a category of solvable potentials, 
we find that these two approaches are indeed related. 

In the next section, we briefly describe supersymmetric quantum 
mechanics (SUSY-QM), and discuss how the constraint of shape invariance 
suffices to determine the spectrum of a shape invariant potential (SIP). 
In sec. 3, we judiciously construct some algebraic operators and show
that the shape invariance constraint can be expressed as an algebraic
condition. 
For a set of shape invariant potentials, we find that the 
shape invariance condition leads to the presence of a SO(2,1) potential
algebra,
 and we thus establish a connection between the two algebraic methods. 
In sec. 4, for completeness, we provide a brief review of SO(2,1)
representation theory. In sec. 5, we derive the spectrum of a class of 
potentials and explicitly show that both methods indeed give identical spectrum.

\newpage
\noindent
{\bf II. SUSY-QM and Shape Invariance}\\

\sss 

A quantum mechanical system specified by a potential $V_{-}(x)$ can 
alternatively be described by its ground state wavefunction $\psi^{(-)}_0$. 
Apart from a constant (chosen suitably to make the ground state energy zero), 
it follows from the Schr\"odinger equation that the potential 
can be written as
$V_{-}(x)=\left({{\psi^{''}_0} \over {\psi_0}}\right)$, where prime denotes 
differentiation with respect to  $x$.  
In SUSY-QM, it is customary to express the system in terms of the 
superpotential 
$W(x)=-\left({{\psi^{'}_0} \over {\psi_0}}\right)$ rather than the potential, 
and the ground state wavefunction is then given by 
${\psi_0} \sim \exp\left( - \int^x_{x_0} W(x) dx \right)$, where $x_0$ is an 
arbitrarily chosen reference point. 
We are using units with $\hbar$ and $2m=1$. 
The Hamiltonian $H_{-}$ can now be written as  
\begin{equation}
H_{-} 
= \left(-{{d^2} \over {dx^2}}+ V_{-}(x) \right)\;=\;
 \left(-{{d^2} \over {dx^2}}+ W^2(x)-{{dW(x)} \over {dx}}\right)\;.
\label{schrodinger}
\end{equation}
However, as we shall see, there is another 
Hamiltonian $H_+$ with potential 
$V_{+}(x)=\left(W^2(x)+{{dW(x)} \over {dx}}\right)$, that is almost 
iso-spectral with the original potential $V_{-}(x)$. In particular, 
the eigenvalues $E^{+}_n$ of $H_{+}(x)$  satisfy $E^{+}_{n}\;=\;E^{-}_{n+1}$,
where $E^{-}_n$ are eigenvalues of $H_{-}(x)$ and $n=0,1,2,\cdots$, i.e. 
except the ground state all other states of $H_-$ are in one-to-one 
correspondence with states of $H_+$. The potentials $V_{-}(x)$ and $V_{+}(x)$ 
are known as supersymmetric partners.  

In analogy with the harmonic oscillator, we now define two operators: 
$A \equiv  \left({d \over {dx}} +W (x) \right)$, and
and its Hermitian conjugate 
$A^{+} \equiv \left(-{d \over {dx}} +W(x)\right)$.
Hamiltonians $H_{-}$ and its superpartner $H_{+}$ are given by 
operators $A^+A$ and 
$AA^+$ respectively.  

Now we shall explicitly establish the 
iso-spectral relationship between states
of $H_{+}$ and $H_{-}$.  
Let us denote the eigenfunctions of $H_{\pm}$ that
correspond to eigenvalues $E^{\pm}_n$, by $\psi^{(\pm)}_n$. 
For  $n=1,2,\cdots~$, 
\begin{eqnarray}
H_{+} \left( A \psi^{(-)}_n \right) &=& AA^{+} \left( A \psi^{(-)}_n
\right)
=A \left( A^{+}A \psi^{(-)}_n \right) =A H_{-} \left( \psi^{(-)}_n
\right)\nonumber \\
&=& E^{-}_n \left( A \psi^{(-)}_n \right)\;.
\end{eqnarray}
Hence, excepting the ground state which obeys $A\psi^{(-)}_0=0$, for any
state $\psi^{(-)}_n$ of $H_-$ there exists a state $A \psi^{(-)}_n$
of $H_+$ with exactly the same energy, i.e.
$E^{+}_{n-1}\;=\;E^{-}_{n}$, where $n=1,2,\cdots$, i.e.
$A \psi^{(-)}_n \propto \psi^{(+)}_{n-1}$.
Conversely, one also has $A^+ \psi^{(+)}_n \propto \psi^{(-)}_{n+1}$.
Thus, if the eigenvalues and the eigenfunctions of $H_{-}$ were
known, one would automatically obtain the eigenvalues and the
eigenfunctions of $H_{+}$, which is in general a completely different 
Hamiltonian.

Now, let us consider the special case where $V_-(x)$ is a SIP. 
This implies that $V_-(x)$ and $V_+(x)$ have the same 
functional form; they only differ in values of other discrete 
parameters and possibly an additive constant. 
To be explicit, let us assume that in addition to the continuous 
variable $x$, the 
potential $V_-(x)$ also depends upon a constant parameter $a_0$; i.e.,
$V_-\equiv V_-(x,a_0)$. 
The ground state of the system of $H_-$ is given by 
${\psi_0(x,a_0)} \sim \exp\left( - \int^x_{x_0} W(x,a_0) dx \right).$
Now, for a shape invariant $V_-(x,a_0)$, one has, 
$V_{+}(x,a_0)=V_{-}(x,a_1)+R(a_0)~,$
where $R(a_0)$ is the additive constant mentioned above. 
Since potentials $V_{+}(x,a_0)$ and $V_{-}(x,a_1)$ differ only by $R(a_0)$,
their common ground state is given by 
${\psi_0(x,a_1)} \sim \exp\left( - \int^x_{x_0} W(x,a_1) dx \right)$.
Now using SUSY-QM algebra, the first excited state of 
$H_-(x,a_0)$ is given by $A^+{(x,a_0)} \psi^{(-)}_0(x,a_1)$. 
Its energy is $E_1^{(-)}$, which is equal to $E_0^{(+)}$. But since 
$E_0^{(-)}=0$, $E_0^{(+)}$ must be $R(a_0)$. Continuing up the ladder
of series of potentials $V_-(x,a_i)$, we can obtain the entire spectrum 
of $H_-$ by algebraic methods of SUSY-QM.
The eigenvalues are given by
$$E^{(-)}_0=0; ~~{\rm and}~~E_n^{(-)}=\sum_{k=0}^{n-1}R(a_k)~~{\rm for}~n>0,$$
and the $n$-th eigenstate is given by 
$$\psi^{(-)}_{n+1}(x,a_0) \sim 
A^+{(a_0)} ~ A^+{(a_1)} \cdots A^+{(a_{n-1})} ~ 
\psi^{(-)}_0(x,a_{n-1})\;.$$
(To avoid notational complexity, we have suppressed the $x$-dependence of 
operators $A(x,a_0)$ and $A^{+}(x,a_0)$.) 

\sss

\noindent
{\bf III. Shape Invariance and Potential Algebra}

\sss

Let us consider the special case of 
a potential $V_-(x,a_0)$ with an additive shape invariance; i.e.
$V_{+}(x,a_0)=V_{-}(x,a_1)+R(a_0)$, where $a_n=a_{n-1}+\delta=a_0+n\delta$, 
where $\delta$ is a constant. Most SIP's fall into this category. 
For the superpotential $W \left(x,a_m \right)\equiv W(x,m)$, 
the shape invariance condition implies 
\begin{equation}
W^2(x,m)+W'(x,m)=W^2(x,m+1)-W'(x,m+1)+R(m)
\end{equation}
   As described in the last section, this constraint suffices to determine
the entire spectrum of the potential $V_-(x,m)$. In this section, we shall 
explore the possible connection of this method with the potential algebra 
discussed by several 
authors 
(\cite{Alhassid,Barut,Englefield0,Englefield,Wu,Tangerman}).

Since for a SIP, the parameter $m$ is changed by a constant amount each 
time as one goes from the potential $V_-(x,m)$ to its superpartner, it is 
natural to ask whether such a task can be formally accomplished by the 
action of a ladder-type operator. 

With that in mind, we first define an operator 
$J_3= - i\,\frac{\partial}{\partial \phi}$, analogous to the 
$z$-component of the angular momentum operator. 
It acts upon functions in the space described by two coordinates
$x \;{\rm and}\; \phi$, and its eigenvalues $m$ play
the role of the parameter of the potential. We also define two more operators, 
$J^-$ and its Hermitian conjugate $J^+$ by
\begin{equation}
J^{\pm} = 
e^{\pm i\,\phi} 
\left[ \pm \frac{\partial}{\partial x} -
W \left(x, -i\,\frac{\partial}{\partial \phi} 
\pm  
\frac{1}{2}
\right) \right]  \;\;\;
\;.
\end{equation}
The factors $e^{\pm i\,\phi}$ in $J^{\pm}$ ensure that they indeed operate 
as ladder operators for the quantum number $m$. 
Operators $J^{\pm}$ are basically of the same form as the $A^\pm$
operators described earlier in sec. 2, except that the parameter $m$ of the 
superpotential is replaced by operators 
$\left( J_3 \pm  \frac{1}{2} \right)$.
%
With explicit computation we find 
\begin{equation}
\left[J_3,J^\pm \right] = \pm J^\pm ~~,
\end{equation}
and hence operators $J^\pm$  
change the eigenvalues of the $J_3$ operator by unity, similar 
to the ladder operators of angular momentum ($SU(2)$). 
Now let us determine the remaining commutator $\left[J^+,J^- \right] $.
The product $J^+J^- $ is given by
\begin{eqnarray}
\label{JpJn}
J^+J^- &=&
e^{i\,\phi} \left[  \frac{\partial}{\partial x} -
W \left(x,J_3+\frac{1}{2}\right) \right] 
e^{-i\,\phi} \left[ -\frac{\partial}{\partial x} -
W \left(x,J_3 -\frac{1}{2} \right) \right] \nonumber\\
&=& \left[ -\frac{\partial^2}{\partial x^2}+ 
W^2\left( x,J_3-\frac{1}{2} \right) - 
W' \left( x,J_3-\frac{1}{2} \right) \right]\; 
\end{eqnarray}
Similarly, 
\begin{equation}
\label{JnJp}
J^-J^+ = \left[ -\frac{\partial^2}{\partial x^2}+ 
W^2 \left( x,J_3+\frac{1}{2}\right) + 
W'\left( x,J_3+\frac{1}{2} \right) \right]\; .
\end{equation}
Hence the commutator of operators $J_+ $ and $J_-$ is given by
\begin{eqnarray}
\label{Jp,Jn}
\left[J^+,J^-\right] 
&=& \left[ -\frac{\partial^2}{\partial x^2}+ 
W^2 \left( x,J_3-\frac{1}{2} \right) -
W' \left( x,J_3-\frac{1}{2} \right) 
\right]\; 
\nonumber \\
& &\;-\;
\left[ -\frac{\partial^2}{\partial x^2}+ 
W^2 \left( x,J_3+\frac{1}{2} \right) +
W' \left( x,J_3+\frac{1}{2} \right) 
\right]\; 
\nonumber \\
&=& -R \left( J_3+\frac{1}{2} \right)\;, 
\end{eqnarray}
where we have used the constraint of shape invariance, i.e.
$V_{-}(x,J_3-\frac{1}{2})-V_{+}(x,J_3+\frac{1}{2})=-R(J_3+\frac{1}{2})$.
Thus, we see that Shape Invariance enables us to close the algebra 
of $J_3$ and $J^\pm$ to 
\begin{equation}
\left[J_3,J^\pm \right] = \pm J^\pm~~,~~~~
\left[J^+,J^-\right] = -R \left( J_3+\frac{1}{2} \right)~~.
\label{algebra}
\end{equation}

Now, if the function $R(J_3)$ were linear in $J_3$, the algebra of 
eq.(\ref{algebra}) would reduce to that of a SO(3) or SO(2,1). 
Several SIP's are of this type,  among them
are the Morse, the Rosen-Morse and the P\"oschl-Teller I and II potentials.
For these potentials, $R\left( J_3+\frac{1}{2} \right)=2~J_3$, and 
eq.(\ref{algebra}) reduces to an SO(2,1) algebra and thus establishes 
the connection between shape invariance and potential algebra. 
Even though there is much similarity between 
SO(2,1)  and SO(3)       algebras, there are 
some important differences between their  representations. 
Hence, for completeness, we will briefly describe the unitary representations 
of SO(2,1) and refer the reader to \cite{Adams} for a more detailed 
presentation.\\

\sss

\noindent
{\bf IV. Unitary Representations of SO(2,1) Algebra}

\sss

In this section, we shall briefly review the SO(2,1) algebra and its unitary 
representations (unireps). This description is primarily based upon a review
article by B.G. Adams, J. Cizeka and J. Paldus (1987). The generators of
the SO(2,1) algebra satisfy 
\begin{equation}
\left[ J_3, J^{\pm} \right] = \pm J^{\pm}\;\;;
\left[ J_+, J_- \right] = -2J_3~~,                               
\label{SO_2_1}
\end{equation}
where 
$J^{\pm}$ are related to their Cartessian counterparts by 
$J^{\pm}=J_1~\pm ~J_2$. (For the familiar SO(3) case, one has 
$\left[ J_+, J_- \right] = + 2J_3$).
The Casimir of the SO(2,1) algebra is 
\begin{equation}
J^2 = - J^+ ~J^- ~+~J_3^2~-J_3~~\; = \; - J^- ~J^+ ~+~J_3^2~+J_3~~.
\label{J_sq}
\end{equation}

In analogy to the representation of angular momentum algebra, one can choose 
$J^2$ and one of the $J_i$'s as two commuting observables.
However, unlike the SO(3) case, each such choice of a pair generates 
a different set of inequivalent representations. For bound states, we 
choose the familiar 
representation space of states $|j,m \rangle$ on which the 
operators $\{J^2,J_3\}$ are diagonal:
$J^2 |j,m \rangle = j(j+1) |j,m \rangle$, 
$J_3 |j,m \rangle =  m |j,m \rangle$.
Operators $J^{\pm}$ act upon $|j,m \rangle$ states as ladder operators:
$J^{\pm} |j,m \rangle = \left[-(j\mp m)(j \pm m+1)\right]^{1\over 2}~ 
|j,m+1 \rangle$.
Since the quantum number $m$ increases in unit steps for a 
given $j$, the general value for $m$ is of the form $m_0+n$, 
where $n$ is an integer and $m_0$ is a real number. 
There is also another constraint on the quantum numbers $m$ and $j$.  
In unitary representations, $J^+$ and $J^-$ are 
Hermitian conjugates of each other, and $J^+ J^-$ and
$J^- J^+$ are therefore positive operators. This implies
$\left[-(j\mp m)(j \pm m+1)\right]
~=~-\left[ \left(j+{1\over 2}\right)^2~-~\left(m+{1\over 2}\right)^2\right]
\geq 0$.
These constraints can be illustrated on a two dimensional planar diagram 
[Fig. 1] depicting the allowed values of $m$ and $j$. Only the open
triangular areas DFB, HEG and the square AEFC are the allowed regions. 
The values of $|m|$ are no longer bounded by $j$, and depending on the $m_0$
(the starting value of $m$), representations multiplets are either 
semi-infinite (bounded from below or above) or completely unbounded.  
Thus there is no finite (nontrivial) unitary representation of SO(2,1).
In general, there are four classes of unireps. \\

\begin{tabular}{l l l}
$D^+(j)$ & 
$
\begin{array}{l}
	{\rm Bounded ~from ~below}\\ 
	(j,m_0) {\rm ~lie ~along} \\
	{\rm ~the ~segment ~AB}
\end{array} 
$
&
$\left\{
   \begin{array}{l}
	m=-j+n; ~~n=0,1,2,\cdots,\\
	j<0,
   \end{array} 
\right.
$ 
\\
&&\\

$D^-(j)$ &
$
\begin{array}{l}
	{\rm Bounded ~from ~above}\\ 
	(j,m_0) {\rm ~lie ~along} \\
	{\rm ~the ~segment ~AG}
\end{array} 
$
&
$\left\{
   \begin{array}{l}
	 m=j+n; ~~n=0,-1,-2,\cdots,\\
	 j<0,
   \end{array} 
\right.
$ 
\\
&&\\
$D_s(j,m_0) $   &  
$   \begin{array}{l}
	(j,m_0)~{\rm  lie ~in }\\
	{\rm ~the ~square~ area} 
   \end{array} 
$
& 
$\left\{
   \begin{array}{l}
	 m=m_0+n; ~~n=0,\pm 1,\pm 2,\cdots, \\
	 j(j+1)<(|m_0|-1)|m_0|; \\
	~-{1\over 2} <m_0 <-{1\over 2} ~,
   \end{array} 
\right. 
$
\\
&&\\
$D_p(j,m_0)$ &  
$   \begin{array}{l}
     {\rm Unbounded ~and} \\{\rm complex} ~j
   \end{array} 
$
& 
$\left\{
       \begin{array}{l}
	m=m_0+n; ~~n=0,\pm 1,\pm 2,\cdots, \\
	-{1\over 2} <m_0 <-{1\over 2} ~,\\
	j=-{1\over 2} + i\beta.
      \end{array}
\right. $
\end{tabular} 

\vspace*{.25in}

Here we will be interested in representations that are bounded from 
either below or above. 
Such representations fall in triangular areas DFB and HEG. 

For the $D^+$ 
representation, the starting value of $m$ can be anywhere on the darkened 
part of the line AB; other allowed values of $m$ are then obtained by the 
action of the ladder operator $J^{+}$.
Owing to the equivalence of $D^+(j)$ and $D^+(-j-1)$, they correspond to 
the same value of $j(j+1)$.
One could have equivalently started anywhere on the segment CD as well 
and used $D^+(-j-1)$. Both are equivalent and each is unique. 
Similarly, for complete $D^-(j)$ ($D^-(-j-1)$) representation, one starts 
from AG (GH) and generates all other states by the action of the $J^-$
operator.

\sss
\noindent
{\bf V. Example}
\sss

As a concrete example, we will examine the Scarf potential which can be
related to the P\"oschl-Teller II potential by a redefinition of the 
independent variable. We will show that the shape invariance of the 
Scarf potential automatically leads to its potential algebra:
SO(2,1). (Exactly similar analysis can be carried out for the Morse, 
the Rosen-Morse, and the P\"oschl-Teller potentials.)
The Scarf potential is described by its superpotential
$W(x,a_0,B)=a_0 {\rm tanh}\,x + B {\rm sech}\,x$. 
The potential 
$V_-(x,a_0,B)=W^2(x,a_0,B)-W'(x,a_0,B)$
is then given by 
\begin{equation}
V_-(x,a_0,B)    = 
\left[ B^2-a_0(a_0+1) \right] {\rm sech}^2\,x + 
B(2 a_0+1) {\rm sech}\,x ~{\rm tanh}\,x +a_0^2~~.
\end{equation}
The eigenvalues of this system are given by (\cite{Cooper})
\begin{equation}
E_n=a_0^2-\left( a_0-n\right)^2~~.
\label{eigenvalue}
\end{equation}
The partner potential $V_+(x,a_0,B)=W^2(x,a_0,B)+W'(x,a_0,B)$ is given by
\begin{eqnarray}
V_+(x,a_0,B)
&=&
\left[ B^2-a_0(a_0-1) \right] {\rm sech}^2\,x + 
B(2 a_0-1) {\rm sech}\,x ~{\rm tanh}\,x +a_0^2~~.
\nonumber\\
&=&
V_-(x,a_1,B) +a_0^2-a_1^2~~, 
\label{algebra-s}
\end{eqnarray}
where  $a_1=a_0-1$.
Thus, $R(a_0)$ for this case is $a_0^2-a_1^2=2a_0-1$, linear in $a_0$.

Now, following the mechanism of the sec. 2, consider a set of operators
$J^\pm$ which is given by 
\begin{equation}
J^{\pm} =                
e^{\pm i\,\phi} 
\left[ \pm \frac{\partial}{\partial x} -
\left\{ \left( -i\,
\frac{\partial}{\partial \phi} \pm \frac{1}{2} \right ) ~{\rm tanh}\,x + 
B ~{\rm sech}\,x  \right\}
\right]~~.
\end{equation}
  Note the similarity between the operators $J^{\pm}$ and
  operators $A^{\pm}$ defined in sec. 2. Since only the parameter $a_0$
  changes in the shape invariance condition, it is replaced by 
  $J_3\pm \frac{1}{2}$.
  Commutators of these operators with $J_3=-i\,\frac{\partial}{\partial \phi}$
can be shown to close on $J^{\pm}$, as discussed in general in Sec. 2.
Now, from eq.(\ref{algebra}) and (\ref{algebra-s}), the commutator of 
$J^{\pm}$ operators is given by $-2 J_3$, thus forming a closed SO(2,1) 
algebra. Moreover, the operator $J^+ J^-$, acting on the basis 
$|j,m \rangle$ gives: 
\begin{eqnarray}
 {J^+ J^-} & \equiv  &
\left[ B^2-\left( m^2-\frac{1}{4} \right) \right] {\rm sech}^2\,x \nonumber \\
& &  + 
B \left( 2 \left( m-\frac{1}{2} \right) + 1 \right) {\rm sech}\,x 
~{\rm tanh}\,x +
\left( m-\frac{1}{2} \right)^2~~.
\end{eqnarray}
which is just the $H_{\rm scarf} \left(x,m-\frac{1}{2},B\right)$, i.e. the Scarf
Hamiltonian with $a_0$ replaced by $m-\frac{1}{2}$. 
Thus the energy eigenvalues of the Hamiltonian will be the same as that of 
the operator ${J^+ J^-}=J_3^2-J_3-J^2$. Hence, 
the energy is given by $E=m^2-m-j(j+1)$.  
Substituting $j=n-m$, one gets
\begin{eqnarray} 
E_n &=& m^2-n-(n-m)^2~~\nonumber\\
    &=& (m-\frac{1}{2})^2~-\left[n-(m-\frac{1}{2}) \right]^2~~.
\end{eqnarray} 
which is the same as eq.(\ref{eigenvalue}), with  $a_0$ replaced by
$\left( m-\frac{1}{2} \right)$.
Thus for this potential, as well as for the other three potentials 
mentioned above, there are  actually an infinite number of potentials 
characterised by all allowed values of the parameter $m$ that correspond
to the same value of $j$ and hence to the same energy $E$. 
~Hence the name ``potential algebra" (\cite{Alhassid,Wu}).

\vspace*{.25in}

\noindent
Conclusion: 
The algebra of Shape Invariance plays an important role in the solvability of 
most exactly solvable problems in quantum mechanics. Their spectrum can be
easily generated simply by algebraic means. Many of these systems also have 
been shown to possess a potential algebra, which provides an alternate algebraic
method to determine the eigenvalues and eigenfunctions. An obvious question is 
whether these are two unrelated algebraic methods or there is a link between 
them. 
For a subset of exactly solvable potentials, those with $R(a_0)$
linear in parameter $a_0$, we have shown the equivalence of 
their shape invariance property to an SO(2,1) potential 
algebra. As a concrete example, we started with the Scarf potential and 
showed explicitly how shape invariance translates into the 
SO(2,1) potential algebra. 
We determined the spectra using the algebra of SO(2,1) and showed them 
to be the same as that obtained from shape invariance.  

However, we only worked with solvable models for which $R(J_3)$ is a linear 
function of $J_3$. There are many systems for which the above is not true. 
Also 
there were new Shape Invariant problems discovered in 1992 (\cite{Barclay})
for which it is not possible to write the potential in closed form. It will be 
interesting to know whether there are potential algebras that describe these
system, and whether they are connected to their Shape Invariance. These are 
open problems and are currently under investigation.

One of us (AG) would like to thank the Physics Department of the 
University of Illinois for warm hospitality. We would also like to thank Dr.
Prsanta Panigrahi for many related discussion.
%
%
 
\newpage
\noindent
FIGURE CAPTION:\\

\sss
\noindent
FIG 1. Two dimension plot showing the allowed region for $m$ and $j$.

\end{document}